\documentclass{appolb}
\usepackage{graphicx}
  \usepackage{amssymb,amsmath,epsfig,bm,pifont}

   \graphicspath{{./figs/}}
\usepackage{color}

   \newcommand{\RedTn}[1]{\textcolor{red}{#1}}
\begin{document}
\title{Emphasizing the different trends of the existing data for the
       $\gamma^*\gamma \to \pi^0$
       transition form factor%
\thanks{Presented by the second author at Light-Cone 2012, July 08-13, 2012,
        Cracow, Poland.}%
}
\author{A.~P.~Bakulev\footnote{deceased}$^{1}$, S.~V.~Mikhailov$^{1}$, A.~V.~Pimikov$^{1,2}$
\address{$^1$ Bogoliubov Laboratory of Theoretical Physics, JINR,
         141980 Dubna, Russia\\
          E-mail:   mikhs@theor.jinr.ru\\
         $^2$ Departamento de F\'{\i}sica Te\'orica -IFIC,
         Universidad de Valencia-CSIC, E-46100 Burjassot
         (Valencia), Spain \\
         E-mail: alexandr.pimikov@uv.es}
\\ \vspace{2mm}
{N.~G.~Stefanis
}
\address{Institut f\"{u}r Theoretische Physik II,
         Ruhr-Universit\"{a}t Bochum,
         D-44780 Bochum, Germany\\
         E-mail: stefanis@tp2.ruhr-uni-bochum.de}
}
\maketitle
\begin{abstract}
The new data on the $\gamma^*\gamma\!\! \to \!\!\pi^0$ transition
form factor of the Belle Collaboration are analyzed in comparison
with those of BaBar (including the older data of CELLO and CLEO)
using an approach based on light-cone sum rules.
Performing a 2-, and a 3-parametric fit to these data,
we found that the Belle and the BaBar data have no overlap at the
$1\sigma$ level.
While the Belle data agree with our predictions, the Babar data are
in conflict with them.
\end{abstract}

\section{Light-cone sum rules for the process
$\mathbf{\gamma^*(Q^2)\gamma(q^2\simeq 0) \to \pi^0}$}
\label{sec:intro}
The validity of the collinear factorization --- the basis of
applications of QCD to hard processes --- was challenged in the
year 2009 by the experimental data measured by the BaBar Collaboration
\cite{BaBar09} for the kinematics
$Q^2 > m_\rho^2,\ q^2\ll m_\rho^2$.
This year, new experimental data by the Belle Collaboration
\cite{Belle12} were presented that do not indicate such a growth and
are grossly compatible with the QCD expectations.
In this presentation, we show how these data and the previous ones by
the CELLO \cite{CELLO91} and CLEO \cite{CLEO98} Collaborations can be
analyzed within the theoretical scheme of light-cone sum rules (LCSR)s
that incorporates contributions from QCD perturbation theory and
higher-twist corrections.
Within QCD, the pion-photon transition form factor (TFF)
$F^{\gamma^{*}\gamma^{*}\pi^0}$ is given by the matrix element
\begin{equation}
 \int\! d^{4}z\,e^{-iq_{1}\cdot z}
  \langle
         \pi^0 (P)\mid T\{j_\mu(z) j_\nu(0)\}\mid 0
  \rangle
=
  i\epsilon_{\mu\nu\alpha\beta}
  q_{1}^{\alpha} q_{2}^{\beta}
  F^{\gamma^{*}\gamma^{*}\pi^0}(Q^2,q^2)\ ,
\label{eq:matrix-element}
\end{equation}
where $j_\mu$ denotes the quark electromagnetic current and
$Q^2\equiv-q_{1}^2 >0$, $q^2\equiv -q_2^2\geq 0$.
For the asymmetric kinematics
$q^2<m_\rho^2$,
one has to include into the calculation the interaction of the
(quasi) real photon at long distances $\sim O(1/\sqrt{q^2})$.
To accomplish this goal, we apply the approach of LCSRs
\cite{BBK89,Kho99} that effectively accounts for the effects of the
long-distance interactions of the real photon by making use of  a
dispersion relation in $q^2$ and applying quark-hadron duality in
the vector channel \cite{BMS02,BMPS11}.
Taking the limit $q^2\rightarrow 0$, we get \cite{Kho99}
\begin{eqnarray}
  Q^2 F^{\gamma^*\gamma\pi^0}\left(Q^2\right)
= \!\!\!\!\!
&&
  \frac{\sqrt{2}}{3}f_\pi
    \left[
          \frac{Q^2}{m_{\rho}^2}
          \int_{x_{0}}^{1}
                          \exp\left(
                                    \frac{m_{\rho}^2-Q^2\bar{x}/x}{M^2}\right)
                                    \bar{\rho}(Q^2,x) \frac{dx}{x}
\right. \nonumber \\
&&
    \left.     + \int_{0}^{x_0}
                         \!\! \bar{\rho}(Q^2,x) \frac{dx}{\bar{x}}
    \right]\, ,
\label{eq:LCSR-FF}
\end{eqnarray}
where $M^2$ is the Borel parameter in the interval 0.7-0.9~GeV$^2$
and the spectral density is given by
$\bar{\rho}(Q^2,x)=(Q^2+s)\rho^\text{pert}(Q^2,s)$
with
\begin{eqnarray}
\label{eq:R_twists}
\rho^\text{pert}(Q^2,s)
=
  \frac{1}{\pi} {\rm Im}F^{\gamma^*\gamma^*\pi^0}
  \left(Q^2,-s-i\varepsilon\right)=
  \rho_\text{tw-2}+
  \rho_\text{tw-4}+
  \rho_\text{tw-6}+\ldots\, .
\end{eqnarray}
The various twist contributions are defined in the form of
convolutions of the corresponding hard parts with the pion
distribution amplitude (DA) of a given twist \cite{Kho99}.
For instance, for the twist-four contribution we use
the effective description \cite{Kho99}
$
 \varphi_{\pi}^{(4)}(x,\mu^2)
\sim
 \delta_\text{tw-4}^2(\mu^2)\,x^2(1-x)^2
$
with
$\delta_\text{tw-4}^2(\mu^2)=0.19\pm 0.04$ GeV$^2$ \cite{BMS02}.
Here we used the abbreviations
$\bar{x}=1-x$,
$s =\bar{x}Q^2/x$,
$x_0 = Q^2/(Q^2+s_0)$,
where $s_0\simeq 1.5~\text{GeV}^2$ is the effective threshold
in the vector channel.
The leading twist-two contribution has the perturbative expansion
($a_s=\alpha_s/(4\pi)$)
\begin{eqnarray}
\label{eq:T_exp}
  F_{\gamma^*\gamma^*\pi^0}^\text{tw-2}
\sim(
  T_\text{LO}  +
 a_s(\mu^2) T_\text{NLO} +
 a_s^2(\mu^2) T_{\text{NNLO}_{\beta_0}}+\ldots)\otimes\varphi_\pi(x,\mu^2)\, .
\end{eqnarray}

The corresponding contributions from (\ref{eq:T_exp}) to the spectral
density (\ref{eq:R_twists}) have been obtained in \cite{MS09}.
For the term $\rho_{\rm NLO}$ we employ the corrected version computed
in \cite{ABOP10}.
The ``bunch'' of the admissible twist-two pion DAs
$\varphi_\pi(x,\mu^2)$
was determined in \cite{BMS01} within the framework of QCD SRs with
nonlocal condensates (NLC)s.
This DA ``bunch'' (shown graphically in Fig.\ \ref{fig:2D} in
the form of a rectangle) can be effectively parameterized in terms of
the first two Gegenbauer coefficients $a_2$ and $a_4$ to
reduce the expansion to
$
 \varphi_{\pi}^{\rm BMS}(x)
=
6x\bar{x}
 \left[
       1 + a_2 C_{2}^{3/2}(2x-1) + a_4 C_{4}^{3/2}(2x-1)
 \right].
$
Our next goal is to extract a pion DA that provides best agreement
with all sets of the existing data by performing a fit procedure of the
Gegenbauer coefficients $a_n$ within the basis of LCSRs,
cf.\ Eq.\ (\ref{eq:LCSR-FF}).
The results for the combined set of data from CELLO\&CLEO\&Belle --
termed CCBe --- will be contrasted to those from the CELLO\&CLEO\&BaBar
set --- called CCBB.

\section{2D-analysis of the combined data fit}
\label{sec:2D}
\begin{figure}[h!]
\centerline{\vspace{-1mm} \includegraphics[width=0.7\textwidth]{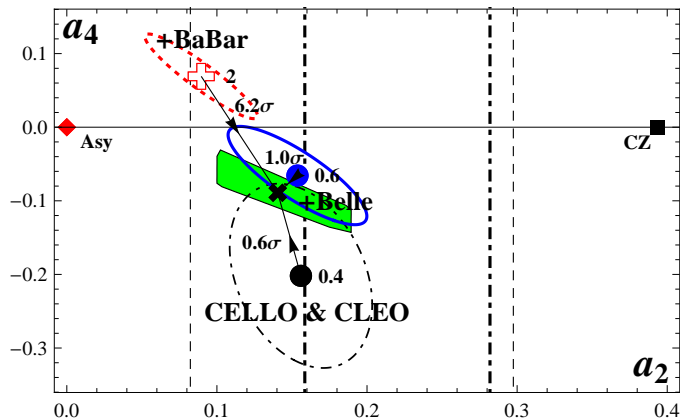}}
\caption{\footnotesize
Error ellipses of various sets of data described in Sec.\ \ref{sec:intro} and
taken at $\mu=2.4$~GeV scale.
The best-fit values of $\chi^2$
are shown at the centers of the ellipses, while differences between data sets
are displayed along the links in units of 1$\sigma$.
The vertical lines show the range of values
computed on the lattice:
dashed line---\protect\cite{Lat06};
dashed-dotted line---\protect\cite{Lat10}.
\label{fig:2D}
}
 \end{figure}

\noindent
In Fig.\ \ref{fig:2D}, we present the confidence region of the
$a_2, a_4$ values in the form of error ellipses in the
$(a_2, a_4)$ plane, obtained by fitting different sets of data.
We take into account only statistical errors and exclude theoretical
uncertainties --- in contrast to our previous work in
\cite{BMS02,BMPS11}.
Using LCSRs with the pion DAs obtained in QCD SR NLC \cite{BMS01},
one arrives at the predictions shown in terms of the (green)
rectangle in comparison with the error ellipses pertaining to the
different sets of data defined in the previous section.
The best-fit values of the $\chi^2$ goodness criterion
$\chi^2_\text{ndf}\equiv\chi^2/\text{ndf}$
(ndf=number of degrees of freedom)
are shown as centers of the ellipses, with the deviations from one data
set to another with reference to the rectangle being displayed along
the links and expressed in units of one standard deviation
(1$\sigma\approx 68\%$).
We conclude that the inclusion of the BaBar data to those obtained
before by CELLO\&CLEO leads to an approximately 3$\sigma$ shift
\cite{PBMS12} of the confidence region away from the (black) ellipse
at the bottom.
The result is represented by the (red) dotted ellipse at the top
accompanied by a significant increase of $\chi^2_\text{ndf}$  from 0.4
to 2 (for the CCBB data set).
If we include to the CELLO\&CLEO data only the Belle data set CCBe,
then the shift of the confidence region is only moderate giving rise to
a slight increase of $\chi^2_\text{ndf}$ from 0.4 to 0.6
(ellipse in the middle).\footnote{Note that we consider the $\chi^2$
goodness criterion for the considered experiments as being the sum of
the individual $\chi^2$ values associated with each experiment.}
We quantify the deviations of these data sets from our theoretical
estimates, encoded in the rectangle, in terms of $\sigma$ values shown
along the links which connect the central data values with the BMS model
(\ding{54}) inside the rectangle.
Here, and in the next figure, the vertical broken lines denote the
constraints from two different lattice simulations: dashed lines
--- \cite{Lat06}; dashed--dotted lines --- \cite{Lat10}.

\begin{figure}[h!]
\centerline{\hspace{-4mm}\includegraphics[width=0.6\textwidth]{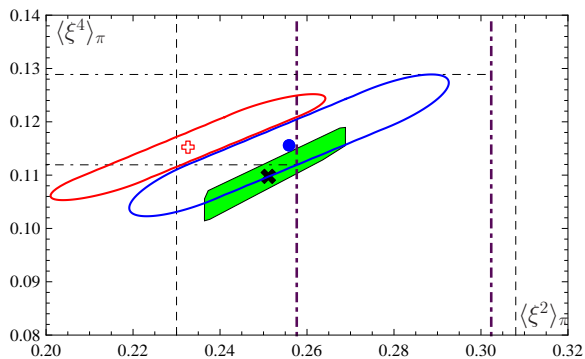}}
\vspace*{-1mm} \caption{\footnotesize (color online).
Predictions for the moments
$\langle \xi^2\rangle$ and
$\langle \xi^4\rangle_\pi$
at the lattice scale $\mu^2_\text{Lat}=4$~GeV$^2$.
The lower stretched ellipse --- solid (blue) line --- corresponds to a
2D fit to the CCBe data, while the upper one (red line) represents the CCBB one.
In contrast to Fig.\ \protect\ref{fig:2D}, we also include here
the uncertainty related to the twist-four coefficient $\delta^2_\text{tw-4}$.
The vertical lines show the range of values
computed in two lattice calculations using the same designations as in Fig.\
\protect\ref{fig:2D}.
\protect\label{fig:Moments}}
\end{figure}

An immediate conclusion from these considerations is that
(i) the BMS DA is inside the 1$\sigma$ CCBe and inside the 0.6$\sigma$
region of the CELLO\&CLEO set, whereas the BMS ``bunch'' greatly
overlaps with both of these error ellipses.
At the same time, the central point of CCBB is 6.2$\sigma$ away and
has no overlap with the CCBe ellipse.
(ii) The asymptotic DA (\RedTn{\ding{117}})
and the Chernyak-Zhitnitsky (CZ) DA ({\footnotesize\ding{110}})
are both more than $6\sigma$ away from CCBe.
(iii) The existing lattice calculations of $a_2$, shown as vertical
lines in Figs.\ \ref{fig:2D},\ref{fig:Moments}, are not restrictive
enough to narrow down the interval of $a_2$, though the narrower band
\cite{Lat10} supports both the CCBe data and our theoretical results
but not the CCBB data.
On the other hand, the previous lattice simulation \cite{Lat06} is
compatible with all sets of data and with our theoretical predictions
as well.

With all these data in our hands, we may attempt to extract values of
the moment
$\langle \xi^4 \rangle_\pi$
by combining the real data from CCBe and CCBB with the results of
lattice simulations.
The combined constraints from CCBe (lower (blue) error ellipse in
Fig.\ \ref{fig:Moments}) and the lattice constraints from \cite{Lat10}
(dashed-dotted vertical lines)
lead to the following predictions for $\langle \xi^4\rangle_\pi$:
$\langle \xi^2 \rangle_\pi \in [0.26 \div 0.30]$
and
$\langle \xi^4 \rangle_\pi \in [0.112 \div 0.13]$.
In contrast, the constraints extracted from the CCBB data (upper (red)
ellipse) do not allow a reliable determination of the
$\langle \xi^4 \rangle_\pi$ range in the background of the lattice
results from \cite{Lat10}.
On the other hand, the constraints from \cite{Lat06} are not stringent
enough to differentiate these sets of data.

\section{3D-analysis of the combined data}
\label{sec:3D}
\begin{figure}[ht!]\noindent
\centerline{
 \includegraphics[width=0.48\textwidth]{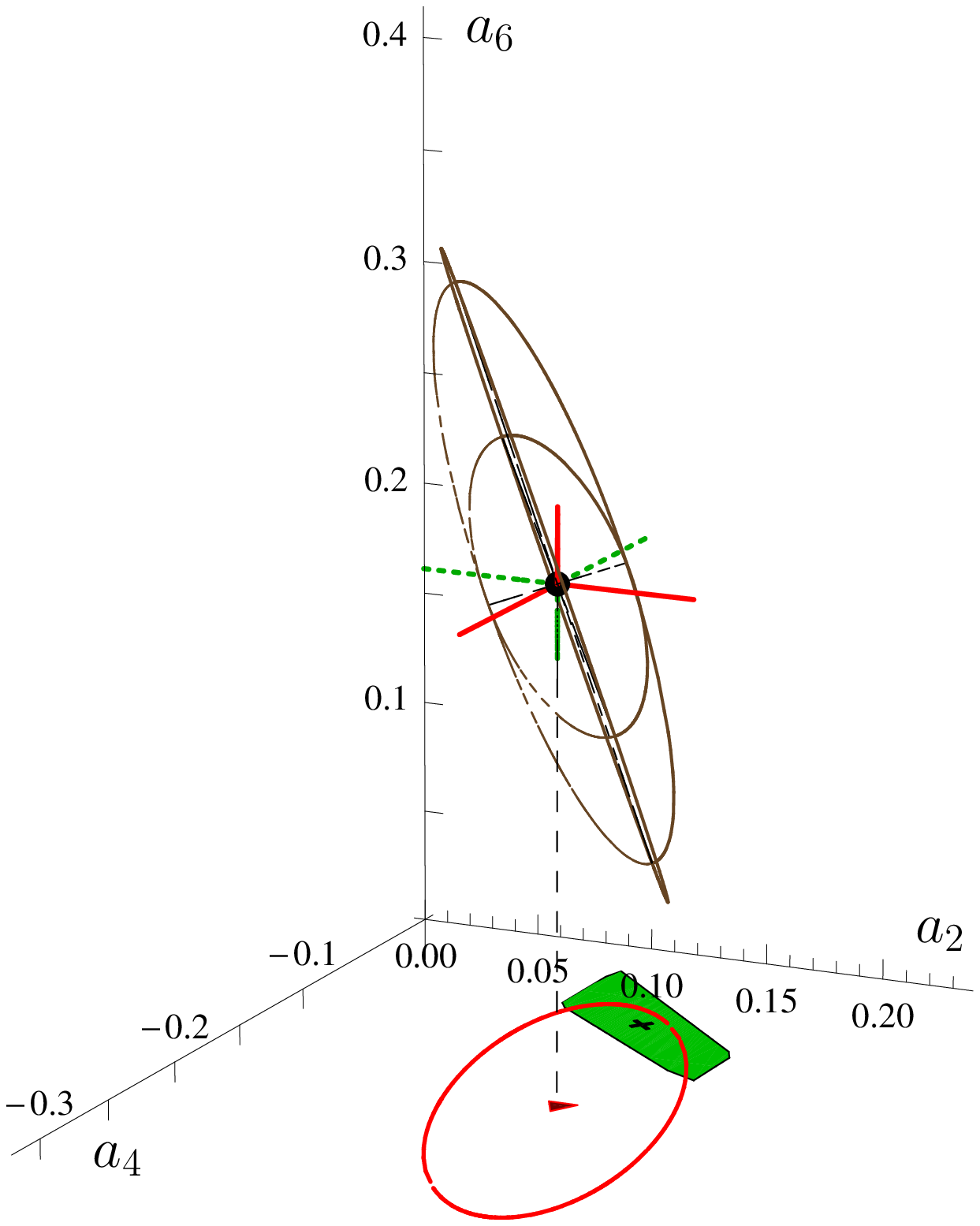}~
  \includegraphics[width=0.48\textwidth]{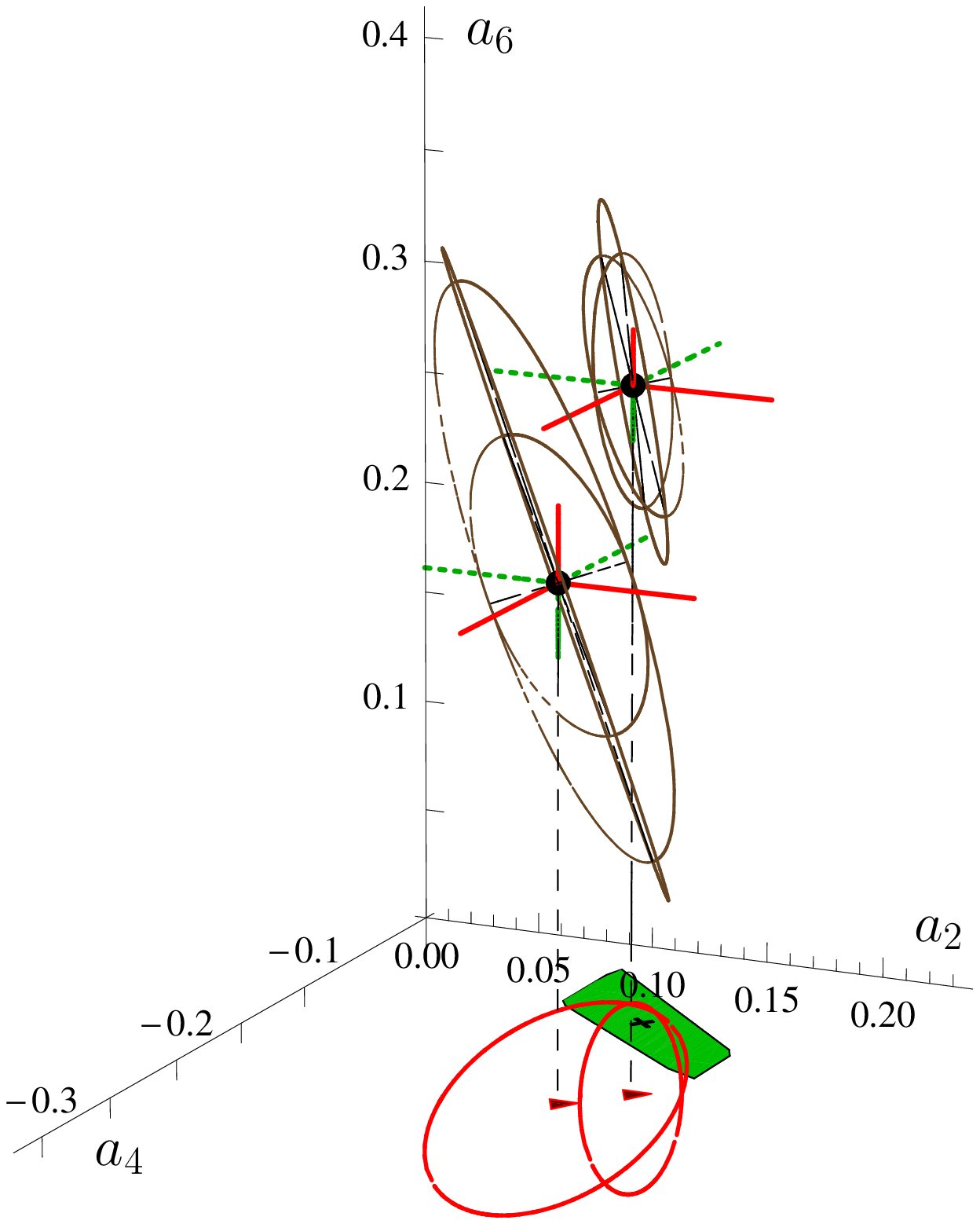}
}
\caption{\footnotesize
3D graphics of the $1\sigma$ error ellipsoids in the space spanned by
the Gegenbauer coefficients $a_n$ $(n=2,4,6)$ for processing the
data on the pion-photon transition form factor within the LCSR
approach using the scale $\mu$=2.4~GeV.
Left. Only CCBe data used.
Right. CCBB fit (smaller ellipsoid) shown in comparison with CCBe data.
More explanations are given in the text.
\label{fig:3D}}
\end{figure}

We step up from a 2D to a 3D analysis of the data, by including the
next coefficient $a_6$.
Then, we get in Fig.\ \ref{fig:3D} fit results in terms of $1\sigma$
ellipsoids with respect to the experimental statistical errors for
the CCBe data set (left panel) and the CCBB (right panel).
The theoretical $\mp \Delta\delta^2_\text{tw-4}$-error is indicated
by a solid (red) cross  for an
increasing value of $\delta^2_\text{tw-4}$, whereas a dashed (green)
cross (closer to the $a_6$ axis) denotes a decreasing value.
The projection of the CCBe $1\sigma$ ellipsoid on the plane $(a_2,a_4)$
is represented, in both panels, by the larger (red) ellipse, while the
smaller one refers to CCBB.
The shaded (green) rectangle encloses the region of $a_2,a_4$
pairs allowed by NLC SRs \protect\cite{BMS01}, with the symbol
({\footnotesize\ding{54}}) marking the BMS pion DA.
All results are shown at the scale $\mu=2.4$~GeV after NLO
evolution.

To further quantify these statements, we supply the ``coordinates'' of
the central point of each of the two displayed ellipsoids in the
following form.
The first number gives the central value of the fit, the next number is
the statistical 1$\sigma$ error, and the third one is  the theoretical
uncertainty due to twist-four.
Then, we have:
CCBe~$(0.157\pm0.057\pm0.056,-0.192\pm0.122\pm0.077,0.226\pm0.161\pm0.033)$
with
~$\chi^2_\text{ndf}=13.1/30$;
CCBB~$(0.177\pm0.054\pm0.056,-0.171\pm0.103\pm0.071,0.307\pm0.096\pm0.024)$
with
~$\chi^2_\text{ndf}=33.3/32$.\\
To conclude: (i) The description of the CCBe data provides a better
$\chi^2_\text{ndf}\approx 0.4$ relative to $\chi^2_\text{ndf} \geq 1$
following from CCBB.
(ii) The CCBe and CCBB ellipsoids are significantly separated from
each other, so that a QCD description of these sets of data requires
substantially different DAs.
(iii) The $(a_2,a_4)$ projections of both ellipsoids have a good
overlap with the BMS ``bunch'', though the CCBe ellipsoid that has no
intersection with it.
An intersection is possible but only at a larger value of confidence level.

\section{Local characteristics of the pion DA}
\label{sec:local}
\begin{figure}[ht]
    \centerline{
  \includegraphics[width=0.46\textwidth]{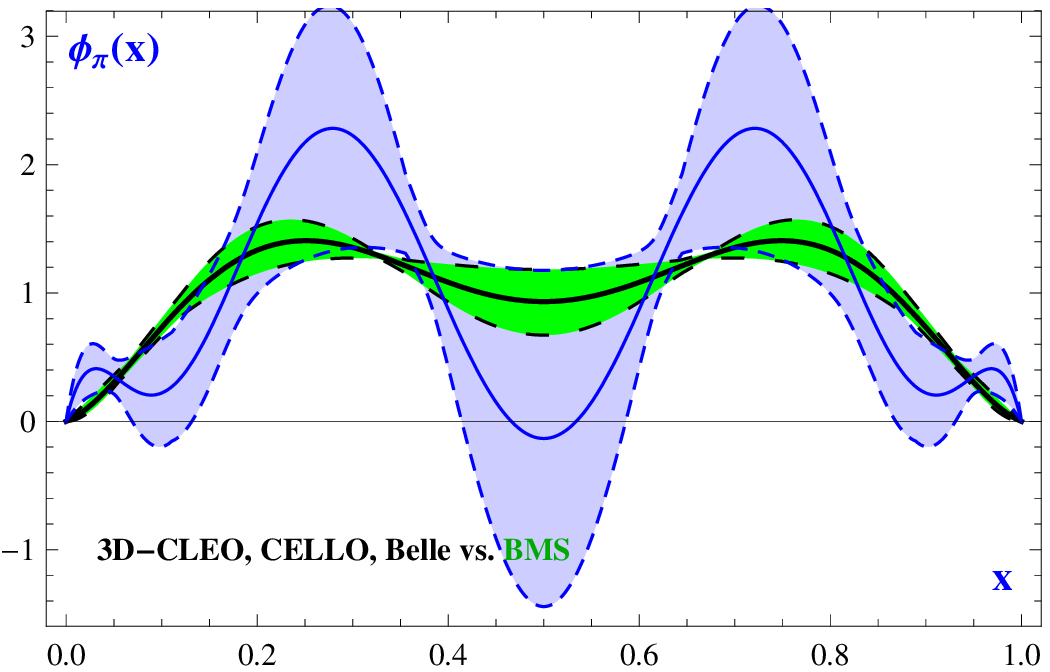}~~
   \includegraphics[width=0.46\textwidth]{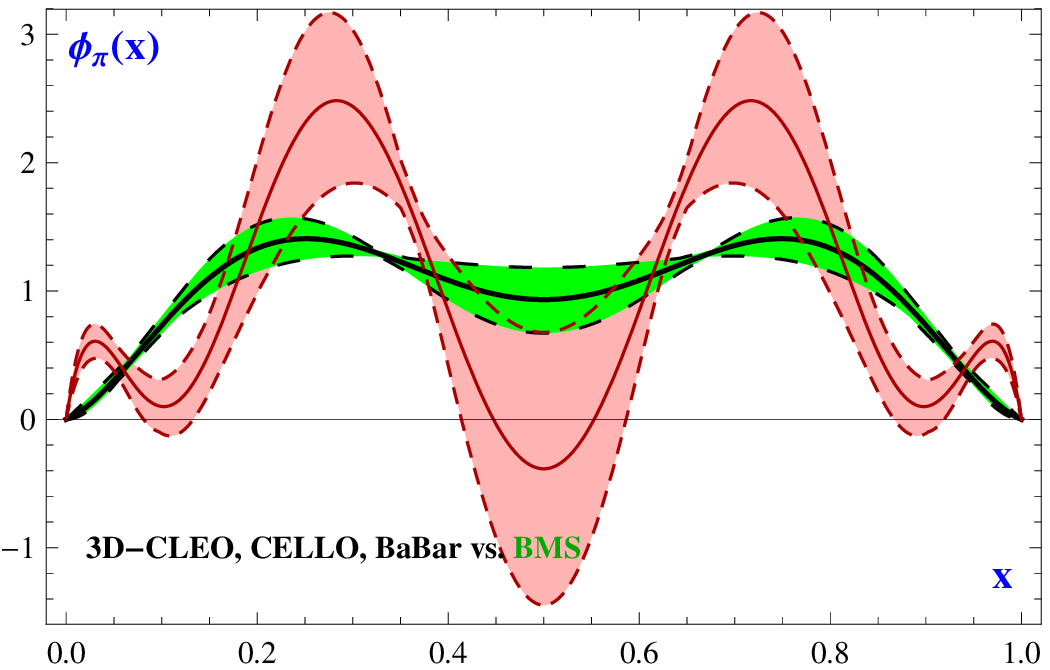}}
\caption{\footnotesize
Left. Comparison of the BMS pion DA ``bunch'' (darker shaded strip in green)
and of the BMS model (black solid line inside this strip) with the 3D
fit to the experimental data on the pion-photon transition form factor.
The solid blue line denotes the best-fit pion DA
(at the scale $\mu=2.4$~GeV, see second paper in \protect\cite{Kho99})
obtained from the analysis of the set CCBe, with the dashed lines indicating
the sum of the statistical errors of the fit and the twist-four uncertainties.
Right. Analogous results obtained with the set CCBB.
\label{fig:piDA}
}
\end{figure}

The confidence region of the coefficients $\{a_n\}$, obtained in
Sec.\ \ref{sec:3D}, can be linked to other characteristics of the pion
DA.
The profiles of the pion DA $\varphi_\pi(x)$, extracted in the 3D fit
procedure, are shown in Fig.\ \ref{fig:piDA}: left panel --- set CCBe;
right panel --- set CCBB.
The BMS ``bunch'' (shaded green strip) and the BMS DA model
(black solid line inside it) are also shown.

\noindent The main difference between the two graphics in Fig.\
\ref{fig:piDA} stems from their distinct behavior near the endpoints
that are concentrated inside the interval $\Delta=0.05$.
Indeed, one observes that the DA (including uncertainties) --- shaded
area in the right panel --- providing best-fit to the CCBB data
deviates significantly from the BMS ``bunch''.
A mathematical tool to quantify the endpoint characteristics of the
pion DA is $D^{(m)}\varphi(\Delta)$, i.e., the average derivative of
$\varphi(x)$ in the interval $\Delta$, that was invented in \cite{MPS10}
and possesses the following properties:
$
\lim\limits_{\Delta\to 0} D^{(m)}\varphi_\pi(0)(\Delta)=\varphi'_\pi(0);
 \lim\limits_{m\to \infty} D^{(m)}\varphi_\pi(0)(\Delta)=\varphi'_\pi(0)
$, while
$
D^{(2)}\varphi(1)= \int_0^1 dx \varphi_\pi(x)/x
$.
The quantity $D^{(2)}\varphi(\Delta)$ can be used \cite{BMPS_QCD2011} to
probe the endpoint behavior of these different sorts of DAs:
\begin{eqnarray}
D^{(2)}\varphi(0.05)=17.2\pm 8.5~\text{ for CCBe};
~25.6\pm 5.25~\text{for CCBB}\, .
\label{eq:2-der}
\end{eqnarray}
We see that the slope in the endpoint region of the
best-fit DA to CCBB is much stronger than for CCBe.
While the CCBe-based DA profile agrees with the BMS ``bunch'',
it is incompatible with the CCBB shape.

\section{Conclusions}
\label{sec:concl}
We presented here an analysis of all experimental data on the
pion-photon transition form factor and compared in detail the results
obtained with two different sets of data: CELLO, CLEO, BaBar ---
(CCBB) vs. CELLO, CLEO, and Belle --- (CCBe).
Our analysis is based on LCSRs at NLO, also taking into account the
twist-four term.
The NNLO$_\beta$ radiative correction was included into the theoretical
uncertainties together with the twist-six contribution \cite{ABOP10}.
The key results can be summarized as follows:
(i) We performed a 2D (parameters $a_2, a_4$) and a 3D
(parameters $a_2, a_4, a_6$) analysis, fitting both the CCBe and the
CCBB data sets.
We found that in both cases the fit to CCBe provides a significantly
better $\chi^2_\text{ndf}$ value.
(ii) The CCBe 2D fit agrees well within error bars with the previous
findings from the CELLO and CLEO fits and the predictions derived from
QCD SR NLC\cite{BMS01} with the BMS ``bunch'' and the BMS model DA.
The fits are also compatible with the constraints on $a_2$ extracted
from lattice computations.
In contrast, the CCBB 2D fit has no overlap with the CCBe result and
it also does not comply with the QCD SR NLC predictions.
(iii) The results of the 3D fit to the CCBe and CCBB data sets do not
intersect at the level of a 1$\sigma$ accuracy.
At the same time the CCBe result is much closer to the BMS ``bunch'',
but still outside the 1$\sigma$ area.
(iv) The qualitative features of the 3D fit to the CCBe and CCBB data
can be differentiated in terms of the behavior of the corresponding
profiles of the pion DAs near the endpoints:
the CCBe has a less pronounced slope and is close to the BMS ``bunch'',
whereas the slope of CCBB is larger and gives no support to the
BMS DAs.

We acknowledge support from the Heisenberg-Landau Program (Grant 2012),
Russian Foundation for Fundamental Research (Grant No.\ 12-02-00613a),
and the BRFBR JINR Cooperation Program (contract No.\ F10D-002).
The work of A.V.P. was supported in part by HadronPhysics2,
Spanish Ministerio de Economia y Competitividad, and EU
FEDER under contracts FPA2010-21750-C02-01, AIC10-D-000598,
GVPrometeo2009/129.

\end{document}